\documentstyle[psfig]{mn}

\newif\ifAMStwofonts

\newcommand{\Teff}{\mbox{$T_{\mbox{\scriptsize eff}}\,$}}

\newcommand{\Msolar}{\mbox{$M_{\odot}\,$}}
\newcommand{\halpha}{\mbox{${\rm H}\alpha$}}
\newcommand{\kms}{\mbox{${\rm km\,s}^{-1}$}}
\newcommand{\vsini}{\mbox{${\rm V}\sin i$}}
\newcommand{\logg}{\mbox{$\log g$}}

\ifoldfss
  \ifCUPmtlplainloaded \else
    \NewTextAlphabet{textbfit} {cmbxti10} {}
    \NewTextAlphabet{textbfss} {cmssbx10} {}
    \NewMathAlphabet{mathbfit} {cmbxti10} {} 
    \NewMathAlphabet{mathbfss} {cmssbx10} {} 
  \fi
  \ifAMStwofonts
    \ifCUPmtlplainloaded \else
      \NewSymbolFont{upmath} {eurm10}
      \NewSymbolFont{AMSa} {msam10}
      \NewMathSymbol{\upi}     {0}{upmath}{19}
      \NewMathSymbol{\umu}     {0}{upmath}{16}
      \NewMathSymbol{\upartial}{0}{upmath}{40}
      \NewMathSymbol{\leqslant}{3}{AMSa}{36}
      \NewMathSymbol{\geqslant}{3}{AMSa}{3E}

       \let\le=\leqslant
       
    \fi
  \fi
\fi 

\ifnfssone
  \newmathalphabet{\mathit}
  \addtoversion{normal}{\mathit}{cmr}{m}{it}
  \addtoversion{bold}{\mathit}{cmr}{bx}{it}
  \newmathalphabet{\mathbfit} 
  \addtoversion{normal}{\mathbfit}{cmr}{bx}{it}
  \addtoversion{bold}{\mathbfit}{cmr}{bx}{it}
  \newmathalphabet{\mathbfss} 
  \addtoversion{normal}{\mathbfss}{cmss}{bx}{n}
  \addtoversion{bold}{\mathbfss}{cmss}{bx}{n}
  \ifAMStwofonts
    \ifCUPmtlplainloaded \else
      \UseAMStwoboldmath
      \makeatletter
      \new@mathgroup\upmath@group
      \define@mathgroup\mv@normal\upmath@group{eur}{m}{n}
      \define@mathgroup\mv@bold\upmath@group{eur}{b}{n}
      \edef\UPM{\hexnumber\upmath@group}
      \new@mathgroup\amsa@group
      \define@mathgroup\mv@normal\amsa@group{msa}{m}{n}
      \define@mathgroup\mv@bold\amsa@group{msa}{m}{n}
      \edef\AMSa{\hexnumber\amsa@group}
      \makeatother
      \mathchardef\upi="0\UPM19
      \mathchardef\umu="0\UPM16
      \mathchardef\upartial="0\UPM40
      \mathchardef\leqslant="3\AMSa36
      \mathchardef\geqslant="3\AMSa3E

       \let\le=\leqslant

    \fi
  \fi
\fi 

\ifnfsstwo
  \DeclareMathAlphabet{\mathbfit}{OT1}{cmr}{bx}{it}
  \SetMathAlphabet\mathbfit{bold}{OT1}{cmr}{bx}{it}
  \DeclareMathAlphabet{\mathbfss}{OT1}{cmss}{bx}{n}
  \SetMathAlphabet\mathbfss{bold}{OT1}{cmss}{bx}{n}
  \ifAMStwofonts
    \ifCUPmtlplainloaded \else
      \DeclareSymbolFont{UPM}{U}{eur}{m}{n}
      \SetSymbolFont{UPM}{bold}{U}{eur}{b}{n}
      \DeclareSymbolFont{AMSa}{U}{msa}{m}{n}
      \DeclareMathSymbol{\upi}{0}{UPM}{"19}
      \DeclareMathSymbol{\umu}{0}{UPM}{"16}
      \DeclareMathSymbol{\upartial}{0}{UPM}{"40}
      \DeclareMathSymbol{\leqslant}{3}{AMSa}{"36}
      \DeclareMathSymbol{\geqslant}{3}{AMSa}{"3E}

       \let\le=\leqslant

    \fi
  \fi
\fi 

\ifCUPmtlplainloaded \else
  \ifAMStwofonts \else 
    \def\upi{\pi}
    \def\umu{\mu}
    \def\upartial{\partial}
  \fi
\fi

\title[Projected rotational velocities of WD\,1614$+$136 and WD\,1353$+$409]
{Projected rotational velocities of WD\,1614$+$136 and WD\,1353$+$409  --
implications for the rate of galactic Type~Ia supernovae.}
\author[P. F. L. Maxted and T. R. Marsh]
       {P. F. L. Maxted and T. R. Marsh \\
        University of Southampton, Department of Physics \& Astronomy,
        Highfield, Southampton, S017 1BJ, UK}
\date{Accepted 1998 March 13.  
      Received 1998 January 26.}

\pagerange{\pageref{firstpage}--\pageref{lastpage}}
\pubyear{1999}

\begin{document}

\maketitle

\label{firstpage}

\begin{abstract}
The white dwarf stars WD\,1614$+$136 and WD\,1353$+$409 are not sufficiently massive
to have formed through single star evolution. However, observations to date
have not yet found any evidence for binarity. It has therefore been suggested
that these stars are the result of a merger. In this paper we place an
upper limit of $\approx 50\kms$ on the projected rotational velocities of both
stars. This suggests that, if these stars are the results of a merger,
efficient angular momentum loss with accompanying mass loss must have occurred.
If the same process occurs following the merging of more massive white dwarf
stars, the predicted rate of Type~Ia supernovae due to merging white dwarfs
may have been greatly over-estimated. Further observations to determine
binarity in WD\,1614$+$136 and WD\,1353$+$409 are therefore encouraged.

\end{abstract}
\begin{keywords}
white dwarfs  -- binaries: close -- stars: rotation -- supernovae: general --
stars: individual: WD\,1614$+$136 -- stars: individual: WD\,1353$+$409 
\end{keywords}

\section{Introduction}
 The finite age of the universe limits the time available for  stars to
evolve. The lowest mass stars have had insufficient time to evolve to the
point where the remnant is seen as a white dwarf star. There is a
monotonically increasing relationship between the initial mass of the star and
the resulting white dwarf star and so this implies a minimum mass for
white dwarf stars. The exact value is uncertain, but is around 0.55\Msolar
(Bragaglia, Renzini \& Bergeron, 1995). Nevertheless, white dwarf stars are
found with masses well below this limit. These white dwarfs are the
consequence of binary star evolution in which the evolution of a star through
the red giant phase is interrupted by a common--envelope phase in which a
companion star is engulfed by the expanding envelope and then rapidly spirals
in towards the core, ejecting the envelope. This arrests the formation of the
degenerate red giant core resulting in an anomalously low mass white dwarf
star. 

 Dramatic evidence for this scenario was provided by Marsh, Dhillon \& Duck
\shortcite{Marsh95}. They observed  7 DA  white dwarfs selected for the low
mass derived for them by Bergeron, Saffer \& Liebert \shortcite{Berg92} from
their spectra. Marsh et~al.\ were able to measure radial velocities with
accuracies of a few \kms\ by using the narrow core of the \halpha\ absorption
line. Periodic radial velocity variations showed at least 5 of the 7 stars to
be binary stars. No evidence for binarity was found for WD\,1614$+$136 or
WD\,1353$+$409.

 It should be emphasized that the observations of Marsh et~al.\ cannot rule
out the possibility that these two white dwarf stars are binaries. Although a
main--sequence companion more massive than 0.1\Msolar\ can be ruled out in both
cases, the presence of another cool white dwarf, very low mass M dwarf or a
brown dwarf, perhaps in a low inclination orbit, cannot be ruled out. With
such strong evidence for the scenario outlined above it would seem to be
inevitable that these white dwarfs were once members of binary systems. The
nature of any companion, or its fate if it is no longer present, remain open
questions.

 The failure of Marsh et~al.\ to detect binarity in WD\,1614$+$136 and
WD\,1353$+$409 led Iben, Tutukov \& Yungleson \shortcite{Iben97} to suggest these
stars are now single stars that are the result of a merger between  a
white dwarf and the companion responsible for the common--envelope phase. In
this paper we show that the detection of a narrow core to the \halpha\ makes
this suggestion very unlikely unless angular momentum loss from the merger
product is extremely efficient.

\section{The rotational velocity of the white dwarfs}
  A merger between a white dwarf and its companion will produce a star which
will, initially at least, have a large angular momentum. If there is no
mechanism to remove this angular momentum, the white dwarf observed now will
be have a high equatorial rotational velocity, $V_{\rm rot}$. This will
usually, though not always, lead to a large projected rotational velocity,
\vsini. This was demonstrated in the case of a merger between two CO
white dwarfs with masses of 0.9\Msolar and 0.6\Msolar by Segretain, Chabrier
\& Mochkovitch \shortcite{Segr97} using a smoothed particle hydrodynamics
simulation. They found that even if 90\% of the angular momentum is lost,
$V_{\rm rot}$ will still be $\sim 1000\kms$.

\begin{figure*}
\psfig{file=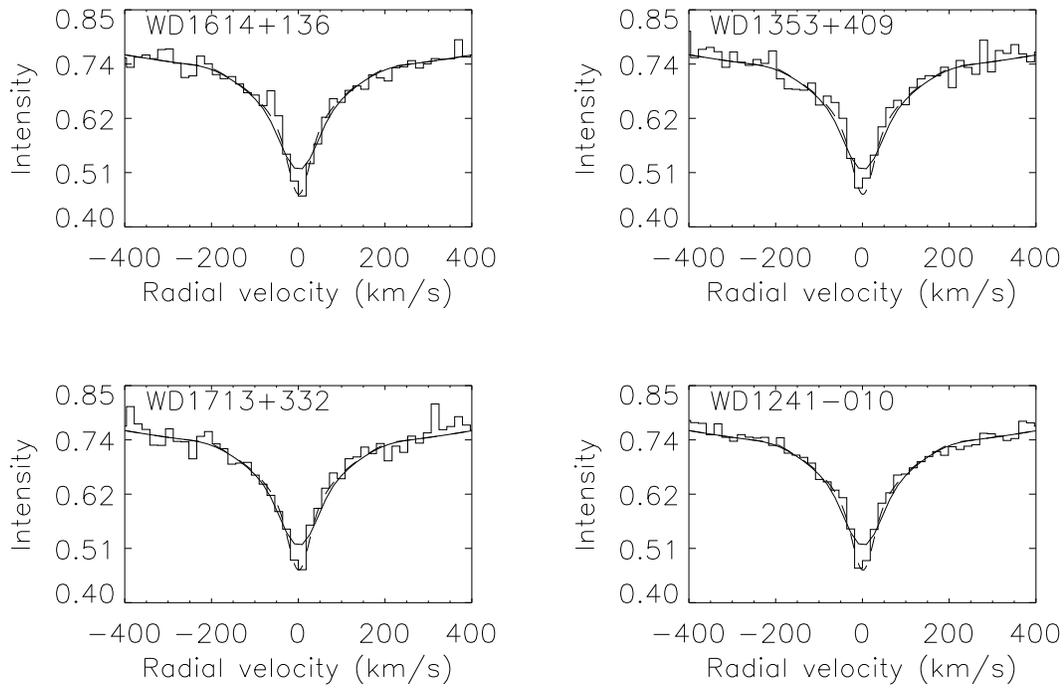}
\caption{\label{SpecFig} Spectra of WD\,1614$+$136, WD\,1353$+$409 and 
the known low-mass binary white dwarfs WD\,1241-010 and 
WD\,1713$+$332 (solid lines, histogram style). Also shown 
are the model spectra of Heber et~al.\shortcite{Heber97} without any rotational 
broadening (smooth, dashed lines) and with an additional 50\kms\ 
rotational broadening (smooth, solid lines).  }
\end{figure*}

 The spectra of WD\,1614$+$136 and WD\,1353$+$409 are shown in Fig.~\ref{SpecFig}.
Also shown are the spectra of the binary white dwarf stars discovered by
Marsh et~al., WD\,1241$-$010 and  WD\,1713$+$332. The contribution of the
companion in these binaries is negligible in this region of the spectrum. The
parameters for these stars derived by Bergeron et~al. and orbital periods if
applicable are shown in Table~\ref{ParTable}. Aside from binarity, these
stars are all quite similar.

 The narrow core of the \halpha\ line is apparent in the spectra of all these
white dwarfs. Also shown in Fig.~\ref{SpecFig} is a model spectrum for
\halpha\ from Heber, Napiwotzki \& Ried \shortcite{Heber97}. This was
calculated using a self-consistent NLTE  model atmosphere for the parameters
\Teff=22\,000K and \logg=8.0. The spectrum for \Teff=25\,000K has a similar
shape and the value of \logg\ has only a minor effect on the shape of the
core. The spectrum has been convolved with a Gaussian profile with full width
at half maximum  of 0.7\AA\ to account for instrumental broadening after the
addition of two broad Gaussian profiles to extrapolate the spectrum into the
wings of the line. The first point to note  is the good match between this
model spectrum and the observed spectra for all four stars. In contrast,
convolving the model spectrum with a rotational broadening profile for \vsini\
of only 50\kms\ results in a very poor match to the observed spectra.
Therefore we adopt an upper limit to \vsini\ for all four stars of 50\kms. 

It is, of course, possible that WD\,1614$+$136 and WD\,1353$+$409 are rapidly
rotating stars which are seen almost pole--on. To obtain some impression of
the probability of this scenario we assumed both stars have the same value of
$V_{\rm rot}$ and then calculated the probability of observing \vsini\ $\le$
50\kms. For a supposed value of $V_{\rm rot}=1000\kms$ the probability is
$1.6\times10^{-6}$. The probability rises to 1/1000 for a $V_{\rm rot} =
200\kms$.

\section{Discussion}
 The low values of \vsini\ found for the white dwarf stars studied here are
not unusual. The study of Heber et~al.\ measured upper limits for 13
white dwarfs from 8\kms to 43\kms. These low rotational velocities are thought
to be due to coupling between the degenerate core of a red giant and its
extended envelope. The narrow core of \halpha\ found for all
the white dwarfs studied by Marsh et~al.\ suggests that this process is not
disrupted by the common envelope phase. It also suggests that very little of
the orbital angular momentum lost by the companion during the common envelope
phase is transfered to the core of the red giant. 

Segretain et~al. discuss the possibility of a red giant phase  following a
merger due to non-explosive burning of the material accreted onto the surface
of the more massive white dwarf. The resulting mass and angular momentum loss
can reduce the mass of the resulting white dwarf below the Chandresekar mass
preventing white dwarf mergers from becoming Type~Ia supernovae. However,
ignition in white dwarf mergers is an extremely complex phenomenon for which
theoretical models have yet to come to a definite conclusion. WD\,1614$+$136
and WD\,1353$+$403 may therefore be key objects for the study of the white dwarf
merging and its implication for the galactic rate of Type~Ia supernovae. If
the rate of galactic Type~Ia supernovae due to merging  white dwarfs is found
to be lower than the observed rate, some other source of Type~Ia supernovae
will need to be found (e.g. supersoft X-ray sources, Branch et~al.\ 1995).

Further radial velocity measuremants may yet show  WD\,1614$+$136 and
WD\,1353$+$403 to be binaries, as may further spectroscopic and photomteric
measurements covering a wide wavelength range, particularly the infrared. Such
observations are to be encouraged given the implications of non-binarity
discussed here. 

\section{Conclusion}
 If WD\,1614$+$136 and WD\,1353$+$409  are now genuinely single stars, their
low projected rotational velocity implies efficient angular momentum loss
following the merging of these stars with the companion star responsible for
the common envelope phase. If the accompanying mass loss occurs following the
merging of more massive white dwarfs, their role as the progenitors of Type~Ia
supernovae is put into doubt.

\begin{table}
\caption{\label{ParTable} Parameters for the white dwarfs 
discussed in this paper.}
\begin{tabular}{lrrr}                     
Name & 
\multicolumn{1}{c}{\Teff/K} &
\multicolumn{1}{c}{\logg} &
\multicolumn{1}{c}{Period/d} \\
WD\,1614$+$136 & 22400 & 7.34 & -- \\
WD\,1353$+$409 & 23600 & 7.54 & -- \\    
WD\,1241$-$010 & 24000 & 7.22 & 3.35 \\
WD\,1713$+$332 & 22000 & 7.40 & 1.13 \\
\end{tabular}
\end{table}

\label{lastpage}
\end{document}